# A metasurface with bidirectional hyperbolic surface modes and position-sensing applications

Chuandeng Hu[1,a)], Xiaoxiao Wu[1,a)], Rui Tong[1], Li Wang[1], YingZhou Huang[2], Shuxia Wang[2], Bo Hou[3,b)], Weijia Wen[1,b)]

[1]*Department of Physics, The Hong Kong University of Science and Technology, Clear Water Bay, Kowloon, Hong Kong, China*

[2]*Department of Applied Physics, Chongqing University, Chongqing 401331, China*

[3]*College of Physics, Optoelectronics and Energy and Collaborative Innovation Center of Suzhou Nano Science and Technology, Soochow University, 1 Shizi Street, Suzhou 215006, China*

[a)] Chuandeng Hu and Xiaoxiao Wu contributed equally to this work.

[b)] Correspondence and requests for materials should be addressed to Weijia Wen (email: phwen@ust.hk) or Bo Hou (email: houbo@suda.edu.cn).

## Abstract

We theoretically and experimentally studied resonance-induced hyperbolic metasurfaces, proving that it is an efficient way to introduce Fano-resonance and decreasethe Q-factor in our system in order to create hyperbolic iso-frequency contours (IFCs) along two orthogonal directions. The metasurface with a continuous topological transition for such IFCs has been designed and experimentally implemented. In particular, two independent self-collimation frequencies corresponding to the transition frequencies in orthogonal directions. As a consequence, we experimentally demonstrated that the metasurface can function as a position- sensor by utilizing the bidirectional hyperbolic surface waves, opening a new avenue for position-sensing.

## Introduction

Surface electromagnetic (EM) waves on two-dimensional interfaces of dissimilar mediums have been a predominant topic of interest among researchers due to their unique features[1-3]. In the optical and near-infrared regime, confined EM waves propagating along the interface between dielectrics and metals are called surface plasmon polaritons (SPPs), which have been widely used in biosensing[4], super-resolution imaging[5,6], and waveguides[7]. However, metals can be regarded as perfect electric conductors (PEC) in low-frequency ranges, e.g., the far-infrared, microwave and terahertz regimes. In these low-frequency regimes, SPPs become very weakly confined Zenneck waves and the evanescent fieldscan extend over several wavelengths[8,9]. Fortunately, designer SPPs or spoof SPPs (SSPPs), which mimic the novel properties of SPPs and overcome the aforementioned limits by using structured PECs were demonstrated in 2004[10]. The metallic/dielectric layers, composed of sub-wavelength and regularly-arranged metallic wires known as metasurfaces, bind EM fields to individual interfaces that exhibit unique anisotropic dispersions that play important roles in manipulating near-field light scattering as well as SPPs and SSPPs[11-13].

Additionally, topology has become significant in many aspects of modern physics. For instance, the Lifshitz transition which describes the transformation of Fermi surfaces of metals from closed to open geometry, and the spin-Hall effect (SHE) of light[14-19] both rely on topology for their analysis. These effects have been predicted for decades; fortunately, the development of metamaterials and metasufaces offer us marvelous platforms to produce them[16,20]. As a result, some topological transport behaviors have been observed in sonic waves and low-frequency EM waves in metasurfaces[21,22], which have been employed to manipulate the SHE of light via an enhanced spin-orbit interaction, and may find potential applications in communications[23]. By using highly anisotropic structures or highly anisotropic media in metasurfaces, hyperbolic dispersion and enhancement in spontaneous emission can be realized by engineering Lifshitz transitions via metamaterials[14]. Very recently, hyperbolic metasurfaces in visible frequency has been demonstrated[24]. Unique properties such as self-collimation due to the special dispersion properties of metasurfaces which were first studied in photonic crystals have attracted immense interest not only in photonic crystals, but also in strong anisotropic SPPs. These properties have led to potential applications in hyperlenses and waveguides[25-28]. Analogously, self-collimating transport exists when the dispersion turns into a flat line, which is a unique shape between the elliptical and hyperbolic IFCs, has been theoretically studied and experimentally measured at microwave frequencies along a single direction[29]. The gradual change in dispersion line shape leads to a topological transition and confines the propagating direction of SSPPs which act as Dyakonov-like waves[30,31]. However, such unique self-collimation only exists in one direction at a single frequency in the previous literature, which limits its application.

Another remarkable resonance responsible for asymmetric spectra, known as Fano-resonance, was proposed long ago and was firstly studied in metasurfaces in 2008, which has brought about numerous applications in electric induced transparency, slow light, lasing, switching and sensing, based on the sharp asymmetrical spectra[32-36]. However, the effects of Fano-resonance on the modes of the metasurface bound below the light line have seldom been studied before.

Position sensors play indispensable roles in microfluidics and biology and plays an essential role in touch screens that are now widely used in daily life. The working principle of recent commercial position sensors comes in four types: capacitive, resistive, surface acoustic waves (SAWs), and infrared (IR)[37-39]. Their technological limitations include complicated calibration processes due to multilayer construction for capacitive and resistive position sensors, restrictions on the device size and substrate materials (piezoelectrics) for SAW sensors, and low sensing resolution for IR sensors[37].

In this paper, we study the effects on the bound modes in metasurfaces due to Fano-resonance and demonstrate that Fano-resonance can lead to hyperbolic IFCs especially for lower Q-factor resonance which acts analogously to symmetrical resonance. We designed a metasurface with Lifshitz-like IFC transitions in different bands in two orthogonal directions corresponding to independent frequencies. Compared with previous work, having independent self-collimation frequencies for orthogonal directions in our design shows great potential for applications not only in superlens and wavefront control, but also in position sensors. As a result, we built a position-sensing prototype device to demonstrate that the designed metasurface can function as a position sensor by working together with recognition systems. The prototype represents a new kind of position sensor based on surface EM waves that has never been reported before.

**Theories and Calculations**

We start with a unit cell composed of a dielectric slab and an anisotropic metallic pattern (H-shape) as shown in Figure 1, in which the dielectric slab is made of nonmagnetic material with relative permittivity 16 and thickness $h$=1 mm. The geometrical parameters of the metallic pattern are chosen with period $D_x$=$D_y$=5.0 mm, $a$=$b$=$c$=3.8 mm, $w_1$=$w_2$=$w_3$=0.4mm and thickness $t$=35 μm. Various EM properties including SSPPs propagation, reflection and transmission, are strongly influenced by both these geometric parameters and the relative permittivity of the dielectric slab and are well predicted by their band structures especially for 2D materials[30,40]. Therefore, we first evaluated the band diagram of our designed metasurface as shown in Figure 2, where Figure 2a represents the band structure below the light line (blue dash lines) with the inset indicating the first Brillouin zone and highly symmetrical points.

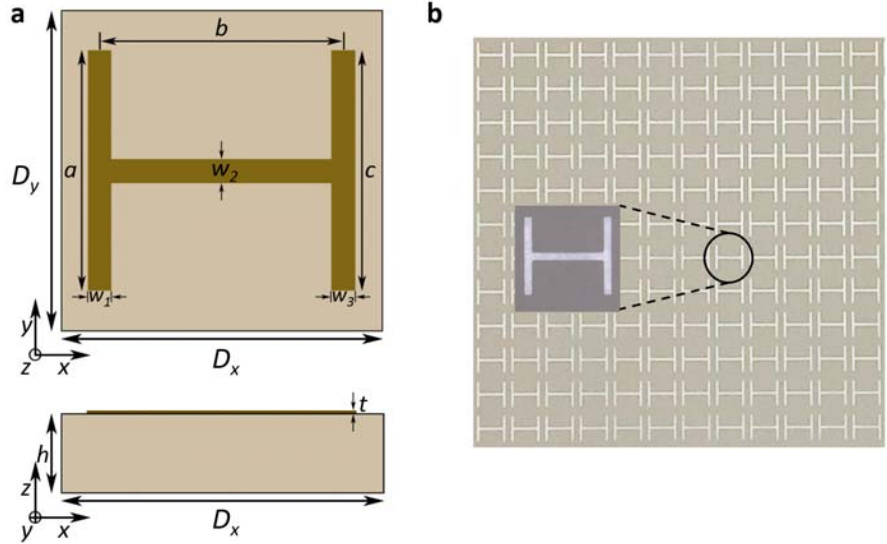

**Figure 1 | a** Geometry of the designed metasurface with $D_x=D_y=5.0$mm, $a=b=c=3.8$mm, $w_1=w_2=w_3=0.4$mm and $t=35$μm. **b** Top-view depiction of a sample with inset showing the details of a single unit.

By further evaluating the IFCs of the first three bands in Figure 2b-d, a topological transition of IFCs under a continuous frequency in the first band has been found, where the shape of the IFCs gradually transforms from a closed ellipse to an open hyperbola and exhibits a flat line at the transition point. The group velocity of SSPP modes is perpendicular to its IFCs, which indicates self-collimation at the transition frequency due to its flat line feature. This phenomenon has been found among other structures, nevertheless, the analytical explanation has not yet been elaborated elaborated[29]. Here we employ effective medium theory (EMT) to interpret the intrinsic physical essence which is necessary in the design of functional metasurfaces.

In most practical cases, the EM response of anisotropic metamaterials can be characterized by the effective permittivity and permeability tensors[41-43]. In our circumstance, permittivity ($\bar{\bar{\varepsilon}}$) is anisotropic with orthogonal principle axes:

$$\overline{\overline{\varepsilon}} = \begin{pmatrix} \varepsilon_{xx} & 0 \\ 0 & \varepsilon_{yy} \end{pmatrix}, \qquad (1)$$

here the coordinates are normalized along the principle axis in order to diagonalize the matrix, and the subscripts *xx* and *yy* denote the x- and y- directions, respectively. By assuming the surface EM waves propagate in the x-y plane and rapidly decay along the z-direction, the dispersion equation of the IFCs in the metasurface can be written generally as[44,45]:

$$\frac{k_x^2}{\varepsilon_{yy}\mu_{zz}} + \frac{k_y^2}{\varepsilon_{xx}\mu_{zz}} = k_0^2, \qquad (2)$$

where $k_x$ and $k_y$ are the propagation vector in x- and y- directions, respectively, $k_0 = \omega/c$ denotes the wave number of the vacuum with angular frequency $\omega$, light velocity in the vacuum *c*, and $\mu_{zz}$ presents the permeability along the z-direction. The dielectric slab is made of nonmagnetic materials, magnetic dipolar moments can only come from the geometrical resonances of the metallic pattern[45], and the multipolar distributions in our designed metasurface are numerically studied (see Supplementary Note 1), which concentrate on electric dipole mode, therefore it is reasonable to set $\mu_{zz} = 1$ as an approximation. Thus, we treat the metasurface as a homogeneous effective-medium slab whose thickness and relative permittivity are to be determined through fitting normal incident transmission spectra[46,47]. By employing Lorentzian dispersion and setting a thickness of 0.1mm (see Supplementary Note 2), the relative permittivity can be expressed as[48]:

$$\varepsilon_{xx} = 163 + \frac{8798.42}{6.1^2 - f^2 - 0.032i \cdot f}$$

$$\varepsilon_{yy} = 175 + \frac{8377.53}{14.12^2 - f^2 - 0.082i \cdot f}$$

(3)

where *f* represents the frequency (in GHz)[49]. The normal incident transmission spectra with electric field polarized along the x- and y- directions are shown in Figure 2d. Here, the red and black points denote spectra from COMSOL Multiphysics simulations, while the solid lines indicate the homogenized approximation. The simulation and approximation are seen to be in good agreement. Equation (3) has been plotted in Figure 2f, where the gray regions (6.1-9.6 GHz and >14.12 GHz) predict hyperbolic IFCs and self-collimation at the transition frequency (6.1GHz). The band diagram from EMT has been studied (see Supplementary Note 3), while the deviation can be understood by considering the strong overlap of frequencies between the different bands and the subwavelength-incompatible geometries at increased the frequency, since EMT become invalid under conditions. Therefore, it does successfully predict the existence of hyperbolic IFCs; namely, hyperbolic IFCs can be introduced by symmetrical resonance and have good accuracy when the structure is in deep subwavelength.

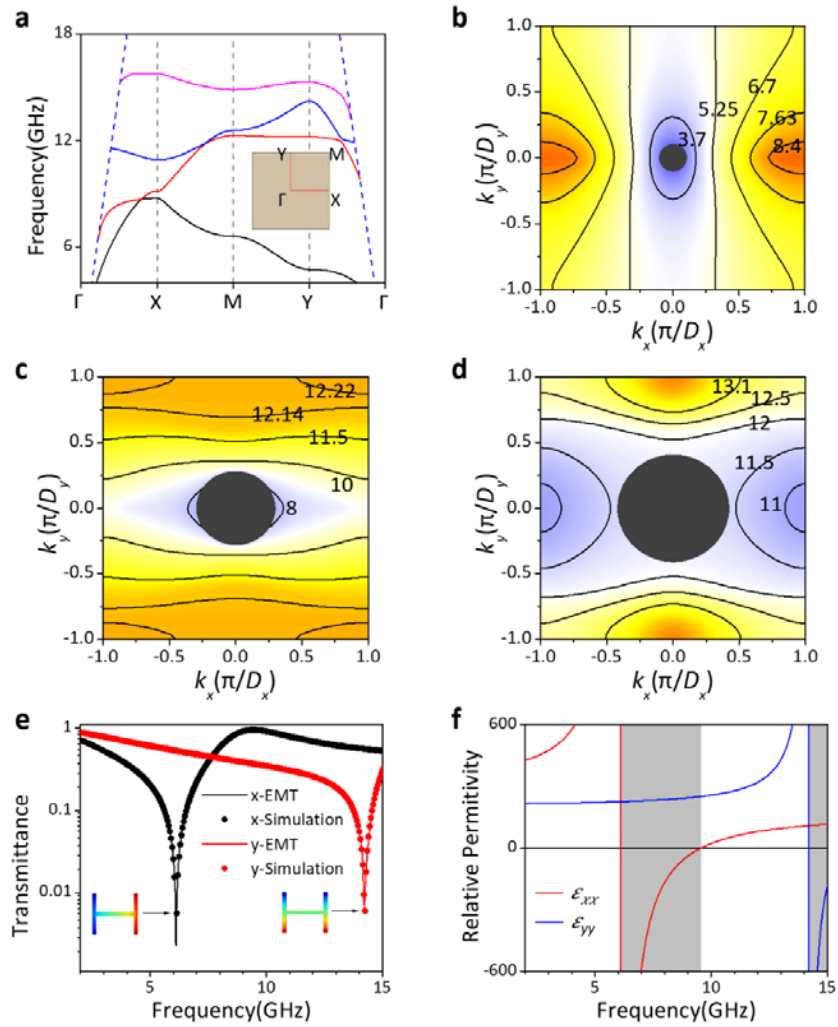

**Figure 2 | Diagrams for *a*=3.8 mm a** Band diagram below light lines with frequency in GHz. The inset are the Brillouin zone and high-symmetrical points. **b-d** IFCs of the first three bands in the first Brillouin, respectively, where the gray discs located at the center are the light cones at 3 GHz, 9 GHz and 12.5 GHz corresponding to b, c and d, respectively, the frequency values are marked in GHz, and the color scheme for frequency increase is from blue to orange. **e** Transmittance spectra at normal incidence obtained from simulations (EMT) when electric field is polarized along the x- and y-directions corresponding to the black line (point) and the red line (point), respectively. Insets are the $E_z$ distribution at different resonance with values variying from –max (blue) to max (red). **f** Relative permittivity obtained from

EMT, where the gray regions indicate frequency ranges of hyperbolic IFCs predicted by EMT.

Unfortunately, the extensive overlap between the frequencies of the second and third bands indicates that there is no independent Lifshitz-like continuous topological transition at this frequency region, while the modes corresponding to the relatively flat IFCs in the second band as shown in Figure 2c are hard to be excited individually. There are two approaches to fix this issue in order to obtain the other independent non-degenerate self-collimation frequency in the y-direction: one is to break the overlap of frequencies between the bands and the other is to introduce a new resonance which can be excited by plane EMWs. First, we numerically calculate the modes at the resonance frequencies as shown in the inset of Figure 2e, where the mode at the higher frequency is a symmetrical dipole mode with dipole momentum along the y-direction. Consequently, there could be a dark mode protected by symmetry along the y-direction. As shown in Figure 3a, when our designed H-shape is symmetrical with respect to the y-direction, the electric quadrupole mode fails to be excited by the incident EM wave. Therefore, our strategy to create non-degenerate self-collimation along the y-direction is to develop Fano-resonance in our system, that is, to sabotage the y-directional symmetry of the metallic pattern. For simplicity, we only decrease the length of *a* and keep other parameters unaltered. The transmission coefficient for this asymmetrical structure when the electric field is polarized along the y-direction can be described though temporal coupled-mode theory[50]:

$$t_n = 1 - \frac{|\alpha_E|^2}{(f - \omega_d) - \frac{|\kappa|^2}{(f - \omega_q)}}, \qquad (4)$$

where $t_n$ denotes the normal incident transmission coefficient, $\alpha_E$ represents the radiative coupling parameter, $\omega_d$ and $\omega_q$ are resonant frequencies corresponding to the quadrupole mode and dipole mode, respectively, and $\kappa$ is the coupling coefficient between these two modes (which is a function of the degree of asymmetry of the pattern) in this normal incidence. For instance, $\kappa$ is equal to 0 when our structure is symmetrical, which indicates only one dip in transmission spectrum due to dipole mode, whereas there are two dips for an asymmetric structure.

Here $a$=3.8 mm, $a$=3.75 mm, $a$=3.7 mm, $a$=3.65 mm, $a$=3.6 mm, $a$=3.55 mm and $a$=3.5 mm are used as of the parameters for the normal incident transmission spectra with the electric field polarized along the y-direction. As shown in Figure 3b, the asymmetric lineshape (Fano line shape), with the symmetry broken along the y-direction, has been directly observed, while the Q-factor can be evaluated by fitting transmission spectra with Fano line shape (see Supplementary Note 4), which decreases exponentially with the decrease in length $a$, as shown in Figure 3c. The resonance frequency shows blue-shift, which is also studied (see Supplementary Note 5). Recently, Fano-resonance with high Q-factor has been used in many areas, such as chemical sensing[51], and slow light[34]. However, it is arduous to establish an efficient form to describe effective permittivity and permeability in this case, and resonances with high Q-factor above the light line will couple with modes below[52]. Therefore, resonances with a lower Q-factor were chosen to achieve our purpose.

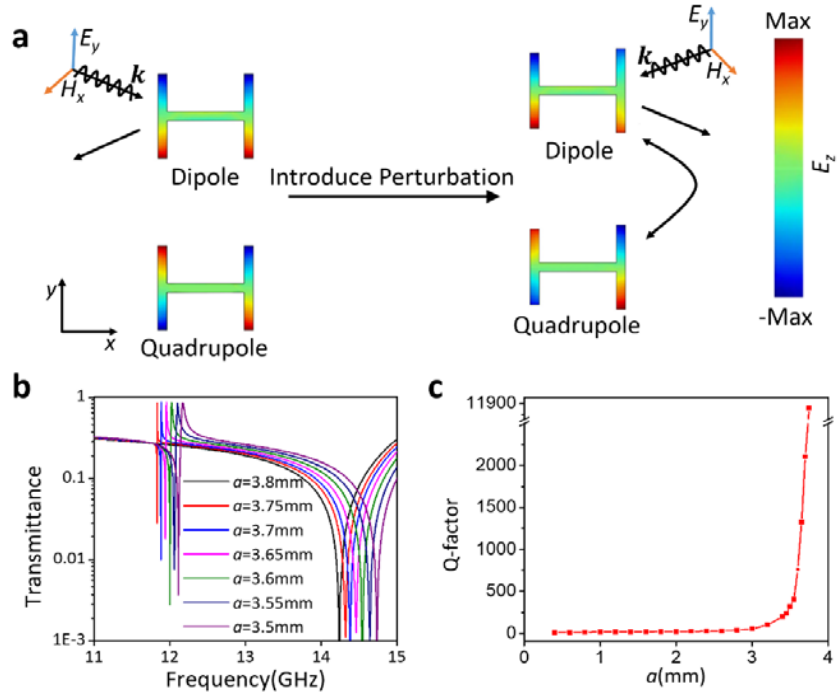

**Figure 3 | a** Fano interference between the electric dipole and quadrupole modes at left and right side indicates the symmetric and asymmetric case, respectively. **b** Normal-incident transmittance spectra with different *a* when electric field is polarized along the y-direction. **c** Variation of Q-factor by fitting Fano line shape with respect to *a*.

The Fano line shape with low Q-factor behaves like symmetrical one, in which the effective parameters can be described via Lorentzian expressions. Here *a*=2.6 mm is picked as an example since the Q-factor is less than 100 as shown in Figure 3c. The contributions to the radiating power in case of *a*=2.6 mm are dominated by the electric quadrupole modes (EQMs), which has been numerically verified (see Supplementary Note 1). The contributions of electric multipole modes (include of EQMs) to effective parameters in metamaterials after homogenization had been studied[53,54]. Particularly, the EQMs in metasurface contribute not only to the effective permittivity in the plane, but also to the effective permeability perpendicular to the plane ($\mu_{zz}$) which is related to such in-plane permittivity[55,56]. And the interested conclusion according to the bianisotropic property predicts that the corresponding resonant frequency for the in-plane permittivity exists when $\mu_{zz}$ vanishes, namely that $\mu_{zz}$ has no contribution on the determination of the shape of IFCs in an application with our case[55]. We note that local constitutive parameters still make sense in the prediction of hyperbolic IFCs even though the metasurface should be regarded as nonlocal once the EQMs are included. It is known that only in-plane effective parameters can be directly determined via normal transmission spectra, therefore the effective permittivity can be fitted, and with a thickness 0.1mm, the expressions read:

$$\varepsilon_{xx} = 163 + \frac{8721.95}{6.68^2 - f^2 - 0.024i \cdot f}$$
$$\varepsilon_{yy} = 175 + \frac{2848.3}{12.65^2 - f^2 - 0.09i \cdot f} + \frac{427.6}{17.2^2 - f^2 - 0.2i \cdot f} \quad (5)$$

Here, the effect of Fano-resonance on $\varepsilon_{xx}$ is neglected because the current distribution is concentrated in the y-direction, which has been numerically verified (see Supplementary Note 6). The transmittance spectra and effective parameters are shown in Figure 4e and 4f, respectively, indicating two hyperbolic IFC regions marked in gray, where $\varepsilon_{xx}$ and $\varepsilon_{yy}$ have opposite signs. To identify our deduction, the IFCs are numerical evaluated shown in Fig. 4b-d, Lifshitz-like topological transition of IFCs exists in the second band with a flat line at the transition frequency, which indicates the other self-collimation frequency with group velocity along the y-direction. The Lifshitz-like transition of the IFCs for the first band is always retained since the physical conditions is consistent, and the self-collimation is very robust while $a$ is being reduced. We note the same phenomenon still exists even when $a$ has been reduced to 0.4 mm so that the metallic pattern can be regarded as 'T-shape'. The IFCs of the first two bands for $a$=1.4 mm and $a$=0.4 mm have been plotted in Figure 5.

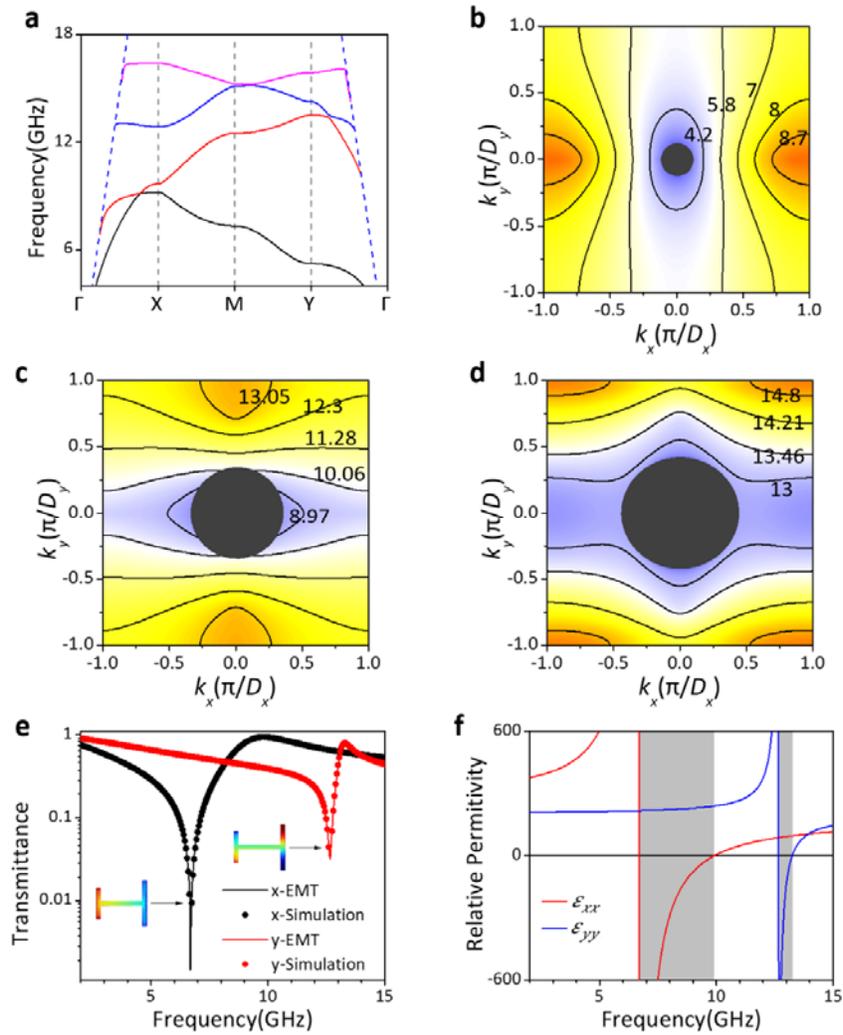

**Figure 4 | Diagrams for *a*=2.6 mm a** Band diagram below light lines with frequency in GHz. **b-d** IFCs of the first three bands in the first Brillouin zone, where the gray discs located on the center are the light cones at 3.7 GHz, 10.5 GHz and 13.2 GHz corresponding to b, c and d, respectively. Frequency values are marked in GHz, and the frequency increases from blue to orange. **e** Transmittance spectra at normal incidence obtained from simulations (EMT) when the electric field is polarized along the x- and y- directions corresponding to the black line (point) and the red line (point), respectively. Insets are the $E_z$ distribution at different

resonances with values varying from –max (blue) to max (red). **f** Effective parameters obtained from EMT, where the gray regions indicate frequency regions of hyperbolic IFCs predicted by EMT.

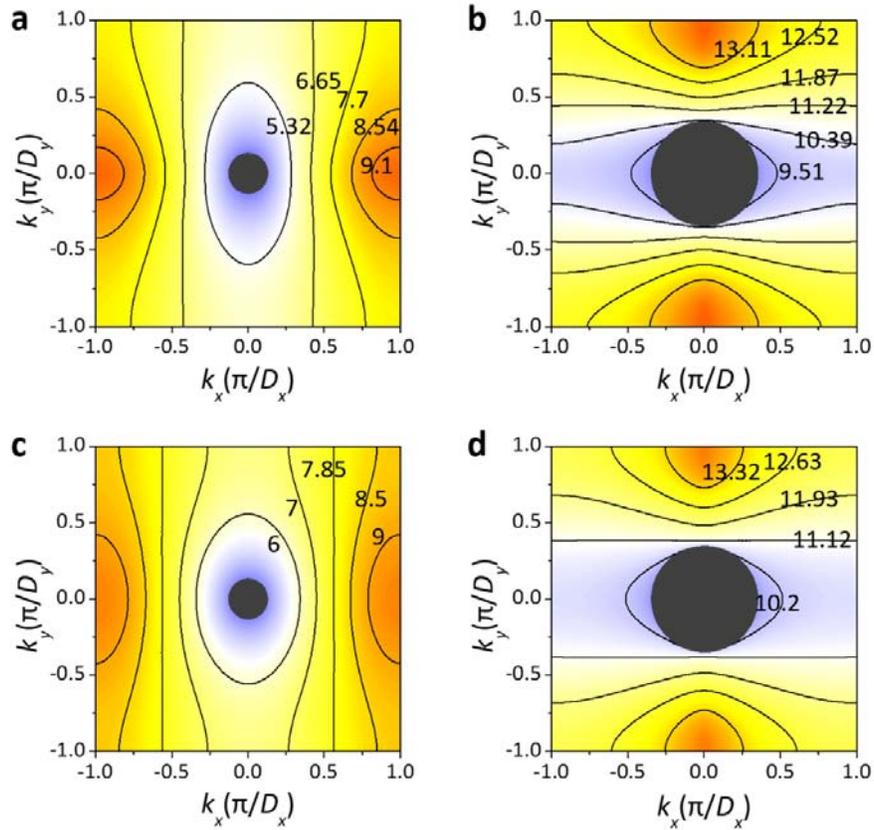

**Figure 5 |** IFCs **a** and **c** are the first bands for *a*=1.4 mm and *a*=0.4 mm, respectively. **b** and **d** are the second bands for *a*=1.4 mm and *a*=0.4 mm, respectively. The gray discs located on the center are light cones at 3.7 GHz for a and c, and 10.8 GHz for b and d, respectively. Frequency values are marked in GHz, and the frequency increase is colored from blue to orange.

To summarize our design, we successfully created a continuous Lifshitz-like topological transition for the IFCs in the y-direction by introducing the low-Q Fano-resonance to our system. Accordingly, after designing continuous non-degenerate topological transitions in orthogonal directions, two flat IFCs with independent frequencies in orthogonal directions can be achieved. Such a phenomenon is very stable under variable geometries. Very recently, the modes corresponding to the transition frequency in a single direction were proven to be diffractionless, soliton-like and sensitive to the surrounding environment[57]. To the best of our knowledge, this is the first time that self-collimation has been demonstrated in both orthogonal directions at independent frequencies which enables many applications especially for position sensors. Meanwhile, recent work on Fano-resonance is mainly concentrated on transmission and reflection spectra, whereas we point out that Fano-resonances are also accountable for the controlling of surface modes located below light line, which opens a new avenuefor manipulating more complicated bands of metasurfaces, which has not been reported before[58-61]. In the next section, we describe the microwave experiment and a sensing prototype further confirming our results.

**Experiment**

A metasurface consisting of 14 x 14 unit cells and excitation units (total size 8 cm x 8 cm) was fabricated. The geometric parameters of the unit cell were chosen as in Figure 1a and the excitation unit was composed of microstrip lines in order to excite surface EM waves (see Supplementary Note 7). The details of our real sample are shown in Figure 1b. The dielectric slab is made of TP1/2 which is a nonmagnetic material with relative permittivity of 16 and tangential loss of 0.001. The metallic patterns were printed in the dielectric slab with 35 μm thick copper and were covered by a negligible thickness of solder on the copper surface in order to prevent oxidation.

The symmetrical case ($a$=3.8 mm) was studied first. Figure 6a shows the distribution of the electric field along the x-direction ($E_x$) which is measured by locating a receiving antenna approximately 1mm above the metasurface when the input frequency is 5.29 GHz along the x-direction. Compared with the IFCs in Figure 2b, the frequency for the flat IFC agrees well with our simulation results, and the slight difference between experiment and simulation is due to the fabrication tolerance and the derivation of relative permittivity. However, we fail to find a self-collimation frequency when excited along the y-direction, and the $E_x$ distribution for relatively unidirectional propagation frequency (11.63 GHz) in our experiment is plotted in Figure 6e. The modes composed of x components have been excited, in which the $E_x$ distribution is observed at a large area instead of along a narrow trail. Compared with Figure 2c and 2d, it is evident that modes in the third band at the same frequency have also been excited, which leads to group velocity along the x-direction.

Next, the asymmetric cases were also studied. Figure 6b and 6f depict the $E_x$ distribution in the case of *a*=2.6 mm when the input frequency is 6.04 GHz along the x-direction and 11.51 GHz along the y-direction. Self-collimation is maintained along the x-direction, which is similar to the symmetric case. However, in contrast to Figure 6e, modes propagation along a narrow region has been observed in Figure 6f, which coincides with the prediction of the IFCs in Figure 4 where the independent flat line exists at the second band. Moreover, the cases for *a*=1.4 mm and *a*=0.4 mm had also been studied, where Figure 6c and 6g corresponds to *a*=1.4 mm and Figure 6d and 6h for *a*=0.4 mm, respectively. Therefore, two independent self-collimation frequencies along orthogonal directions in our designed metasurface have been found, proving that hyperbolic IFCs can be introduced by the low-Q Fano-resonance. The $E_x$ distributions corresponding to the self-collimation frequencies were also studied via COMSOL Multiphysics (see Supplementary Note 8). The self-collimation modes are sensitive to the surrounding environment and the electric field is well confined in a narrow region at the self-collimation frequency. Based on the physical features, we performed the following experiment to demonstrate that self-collimation modes can be used to read the minor changes in the environment, which is the foundation of position sensors.

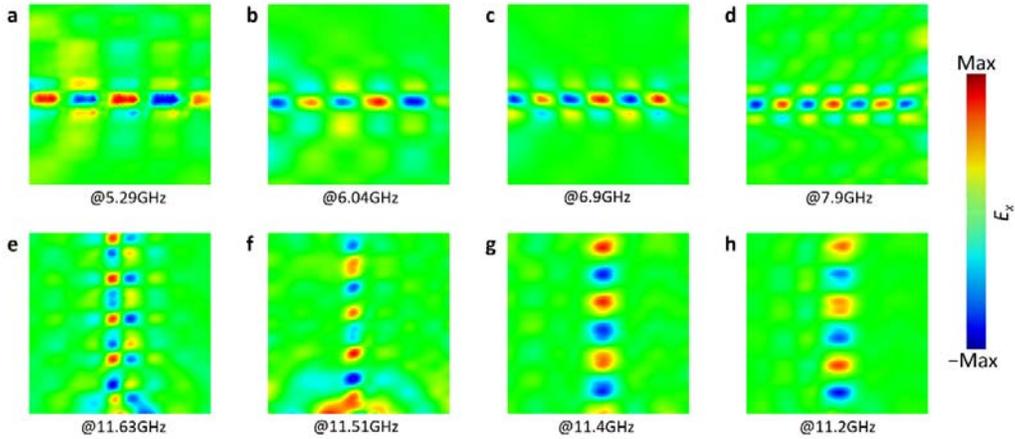

**Figure 6 | a-d** $E_x$ distribution when the metasurface is excited along the x-direction, with $a$=3.8 mm, 2.6 mm, 1.4 mm and 0.4 mm, respectively. **e-h** $E_x$ distribution when excited along the y-direction, with $a$=3.8 mm, 2.6 mm, 1.4 mm and 0.4 mm, respectively.

To change the environment quantitatively, the dielectric cylinders with both radius and height equal to 5 mm were employed, as shown in the inset of Figure 7a (the dielectric cylinders have relative permittivities of 16.5, 20.5, 36, 45 and 69 from left to right). Here, the metasurface of $a$=1.4 mm with excitation along the x-direction is shown in Figure 7a and 7b (the former has the cylinder placed oat the center of the propagation trail and the latter has cylinders placed on the sides of the trail). The S21 amplitude was measured for 6.9 GHz along the x-direction and 11.4 GHz along the y-direction as shown in Figure 7c and 7d, respectively, in the case of $a$=1.4 mm. With $a$=0.4 mm, the cases with 7.9 GHz along the x-direction and 11.2 GHz along the y-direction are plotted in Figure 7e and 7f, respectively. The S21 amplitude was considerably reduced when a dielectric cylinder was put on the propagation trail and slightly fluctuated in the off-trail case, which reveals that interference from the environment of the propagation trail can be registered via S21 amplitude.

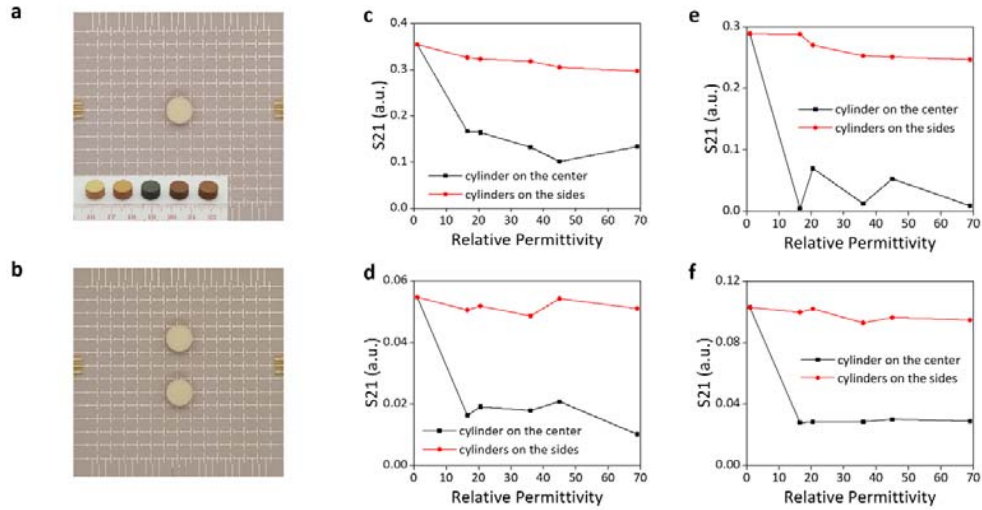

**Figure 7 | Schematic with cylinders a** put on the propagation path with inset showing cylinders of relative permittivities of 16.5, 20.5, 36, 45 and 69 from left to right. **b** on the sides of the propagation path. S21 amplitude corresponding to geometry **c** $a$=1.4 mm, filed along the x-direction. **d** $a$=1.4 mm, field along the y-direction. **e** $a$=0.4 mm, field along the x-direction. **f** $a$=0.4 mm, field along the y-direction.

## Position-Sensing Prototype

The prototype device is composed of two independent microwave sources, four microwave switches, two amplifiers, two detectors, a voltage collector, some coaxial cables and an analysis system, as shown in Figure 8. The switches are controlled by the program so that the ports with the numbers marked on the sample can be periodically opened or closed. For instance, suppose $T$ is the opening time of each switch, with only the four ports labelled 1 open in the time period 0-*T*, only the four ports labelled 2 open in the time period *T*-2*T*, and so on. Then, EM waves emitted from the microwave source can periodically scan from port 1 to port 5 with a periodicity of 5*T*. The EM waves received from switch 2 and switch 4 are amplified because of the limited coupling efficiency from plane EM waves to surface EM waves, and there is substantial loss when transmitting though coaxial cables and switches, so that the power of EM waves can decay to less than the detection limit of the detectors. The EM wave signal is converted into an AC voltage signal when reaching the detector after amplification and collected by the voltage collector. As a result, every row and column is marked by independent numbers, and the 25 total points can be identified and distinguished via the recognition system, when the amplitude of the signal is decreased by the touch of a finger on the surface. In our recognition system, touch location is mapped on the screen and refreshed each period.

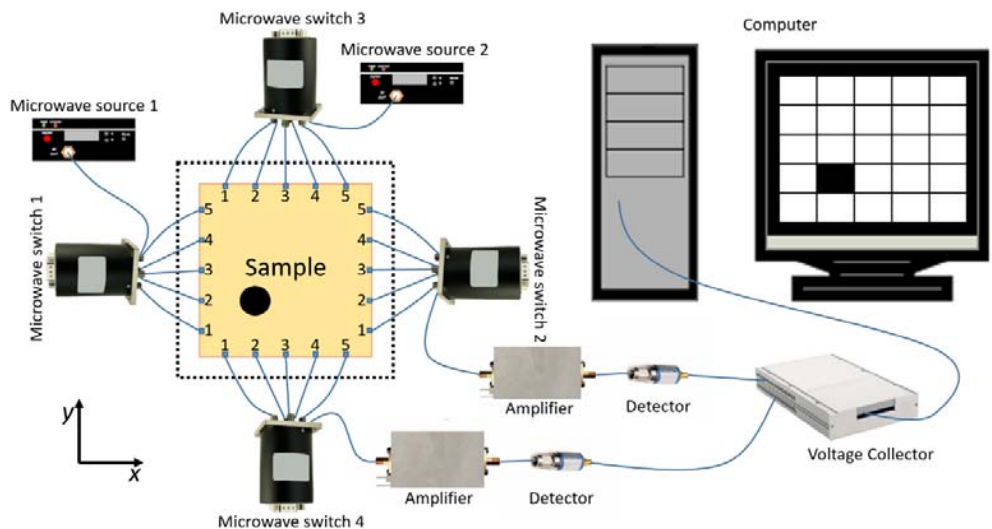

**Figure 8 | Schematic diagram of the position-sensing prototype, where the black disk on the sample represents the perturbation of the environment and the coordinates of the**

**position of the perturbation are analyzed on the computer screen (seeing a real-time demo in Supplementary Video).**

Since the self-collimations are very stable under constant physical state, it is feasible to adjust the geometrical parameters to $D_x=D_y=6.0$ mm, $a=3$ mm, $b=c=5$ mm, $w_1=w_2=w_3=0.4$ mm $d=1$ mm and $t=35$ μm in order to decrease the self-collimation frequency along the x-direction due to the limit of our microwave source. The IFCs of the first two bands under such geometry have been evaluated and plotted (see Supplementary Note 9). The real sample is composed of 20 x 20 units and the excitations were kept unchanged, in which two independent self-collimation frequencies are identified equal to 4.78 GHz and 9.5 GHz corresponding to the x- and y-directions, respectively (see Supplementary Note 9). The open time $T$ is set as 0.6 s due to the limitation of the switching time of our microwave switch, in which the periodicity is 3 s.

In Figure 8, the black disk denotes the perturbation of the environment at the touch of a finger, as shown in the video (see Supplementary video). The coordinates of the position we touched were successfully mapped in the screen which establishes that the metasurface based on the given design can function as position sensors. To the best of our knowledge, this is the first time that a position sensor based on surface EM waves has been realized. Many advantages can be found compared to recent position sensors; for instance, simpler layers and more easily adjusted resolution compared to capacitive and resistive sensors; broader optional substrates and wider application ranges compared to surface acoustic waves sensors[37-39].

**Conclusion and discussion**

In conclusion, we have shown that the Fano-resonance (especially with lower Q-factor) behaves like symmetrical-shaped resonance and can give rise to hyperbolic IFCs in both effective medium theory and numerical simulation. Then, we created topological transition of IFCs in orthogonal directions in different bands by decreasing the Q-factor of the Fano-resonance in our system. Thus, the self-collimating phenomena in both orthogonal directions at transition frequency are verified both numerically and experimentally. Moreover, a position sensor based on surface EM waves was consequently prototyped and demonstrated. This demonstration was enabled by the fact that the self-collimated surface modes have great sensitivity to environmental changes in the propagation trails, and it illustrates one of several promising applications.

**Acknowledgements**

Chuandeng Hu would like to thank Dr. Anan Fang and Dr. Xiao Xiao for useful discussions. The work is supported by an Areas of Excellence Scheme grant (AOE/P-02/12) from Research Grant Council (RGC) of Hong Kong, the Special Fund for Agro-scientific Research in the Public Interest from Ministry of Agriculture of the Peoples' Republic of China (No. 201303045), and the grants from National Natural Science Foundation of China (NSFC) (Nos. 11474212), and the Priority Academic Program Development (PAPD) of Jiangsu Higher Education Institutions.

**Author Contributions**

Chuandeng Hu, Xiaoxiao Wu and W. Wen conceived the original idea. Chuandeng Hu and Xiaoxiao Wu performed finite element simulations. Chuandeng Hu and Bo Hou derived the theory and wrote the manuscript. Yingzhou Huang supported the fabrication process of the sample. Chuandeng Hu carried out the experiments. Rui Tong performed the controlling program of microwave switch. All authors contributed to scientific discussions of the manuscript. W. Wen and Bo Hou supervised the project.

**Methods**

**Simulations**

Throughout this paper, all full-wave simulations are performed via commercial finite element method software COMSOL Multiphysics. The fitting curves are performed though Matlab and Mathematica.

**Position-Sensing Demonstration**

The controlling program of microwave switch is performed via Python, and the display is performed through Labview.

# Additional Information

Supplementary information is available in the online version of the paper.

# Competing Financial Interests

The authors declare no competing financial interests.